\documentstyle[11pt,aaspp4]{article} 
\begin{document}

\title{Wide band observations of the new X-ray burster
SAX~J1747.0-2853 during the March 1998 outburst }

\author{L. Natalucci\altaffilmark{1}, A. Bazzano, M. Cocchi, and P. Ubertini}
          \affil{Istituto di Astrofisica Spaziale {\em(IAS/CNR)},
           via Fosso del Cavaliere, 00133 Roma, Italy}
\author{J. Heise, E. Kuulkers\altaffilmark{2} and  J.J.M. in 't Zand} 
          \affil{Space Research Organization Netherlands {\em (SRON)},
           Sorbonnelaan 2, 3584 CA Utrecht, The Netherlands}

\altaffiltext{1}{e-mail address: lorenzo@ias.rm.cnr.it}
\altaffiltext{2}{Also: Astronomical Institute, Utrecht University, P.O.\ Box 80000,
3507 TA, Utrecht, The Netherlands}

% ------------------- ABSTRACT  ------------------------------------ 

\begin{abstract}

We report on our discovery and follow-up observations of the
X-ray source SAX~J1747.0-2853  
detected in outburst on 1998, March~10 with the 
{\em BeppoSAX} Wide Field Cameras in the energy range 2-28~keV. The source is
located about half degree off the Galactic Nucleus. A total of
14 type~I X-ray bursts were detected in Spring 1998, thus identifying the 
object as a likely low-mass X-ray binary harboring a weakly magnetized neutron
star. Evidence for photospheric radius expansion
is present in at least one of the observed bursts, leading
to an estimate of the source distance of $\sim$~9~kpc.
We performed a follow-up target of opportunity observation with the
{\em BeppoSAX} Narrow Field Instruments on March 23 for a total 
elapsed time of $7.2\times10^{4}$s. 
The source persistent luminosity was 
$2.6\times10^{36}$~erg~s$^{-1}$ in the 2-10 keV energy range.
The wide band spectral data
(1-200 keV) are consistent with a remarkable hard X-ray spectrum detected up
to $\sim150$~keV, highly absorbed at low energies
(${N}_{H}$~$\simeq$~10$^{23}$~cm$^{-2}$) and with clear evidence for an 
absorption edge at $\sim7$~keV.
A soft thermal component is also observed, which can be described by
single temperature blackbody emission at $\sim0.6$~keV.

\end{abstract}


% -------- Section 1: INTRODUCTION AND OBSERVATIONS --------------------- 
\section{INTRODUCTION AND OBSERVATIONS}

The transient bursting source SAX~J1747.0-2853
was discovered during a {\em BeppoSAX} observation of the Galactic Bulge
with the Wide Field Cameras (WFC, \cite{Jag97})
between 1998 March 20.3 and 22.1 (\cite{Zan98}), at $0.3\arcmin$ 
from the centroid position of the X-ray transient GX~+0.2-0.2 observed in 
outburst on 1976 (\cite{Pro78}).

One X-ray burst with a peak intensity of about 1 Crab (2-9 keV) was 
firstly detected on March 21.51.
The source was then observed
with the Narrow Field Instruments (hereafter NFI, see \cite{Boe97} and
references therein) on  board {\em BeppoSAX} on
March 23 (\cite{Baz98}) allowing for an improved positioning
($\alpha=17^{h}47^{m}02^{s}$ and $\delta=-28\arcdeg 52.5\arcmin$,
equinox 2000.0, $1.0\arcmin$ error radius)
and a wide band spectrum to be obtained.
Repeated bursting activity from SAX~J1747.0-2853 has been detected 
during the following WFC observations of April 3-4. 
On 1998 April 13-15, during a pointed NFI observation centred on 1E1743.1-2843
the new transient source was at $13\arcmin$ off-axis and imaged with the 
Medium Energy Proportional Counter (MECS), with a reported flux of
$\sim2\times10^{-11}$~erg~cm$^{-2}$~s$^{-1}$ in the 2-10~keV band 
(\cite{Sid98}). This is consistent with a decreasing of about a
factor of 10 compared to the first NFI observation and with
an outburst exponential decay having an e-folding time of $\sim8$ days. 
A type~I X-ray burst was also observed on April 15 (\cite{Sid98}).

The source re-appeared in outburst
on 2000, March~2 with an average intensity of
$\approx42$~mCrab in the 2-10~keV band, increasing
up to 140 mCrab on March~7.9 as observed by the {\em RXTE}~PCA 
(\cite{Mar00}) and {\em BeppoSAX} NFI (\cite{Cam00}).

% ------------------- Section 2: DATA ANALYSIS -------------------- 
\section{DATA ANALYSIS}

On 1998, March 23 SAX~J1747.0-2853 was strongly detected 
in all {\em BeppoSAX} NFI instruments. 
We evaluated the source intensity in each energy channel of the 
concentrators 
by extracting the counts in a circular region of
$4\arcmin$ radius and scaling for the
corresponding effective area. In the Field-Of-View (FOV) of the MECS
and of the Low Energy
Concentrator Spectrometer (LECS) two additional X-ray 
sources, coincident with 1E~1743.1-2843 and the Sgr-A complex are 
clearly detected, along with the diffuse emission component 
which is present in the vicinity of Sgr-A.
In order to properly subtract the background,
we extracted the counts from the MECS and LECS from an annulus
at a distance of $\sim6\arcmin$ from the centre of the FOV
(thus avoiding the two contaminating point sources) 
having the same geometric area of the source circle. We then applied the 
appropriate correction for effective area. The background 
intensity is low 
($\leq2$~\% respect to the source) and we estimate that the overall
systematic error in the net source intensity 
is not larger than $\sim1$\%.  

The data reduction of the non-imaging instruments, 
namely the Phoswich Detector System (PDS) and the
High Scintillation Proportional Counter (HPGSPC) with an aperture of
1.5~$\arcdeg$ and 1.4~$\arcdeg$ respectively,  
is affected by confusion with nearby sources. 
In order to assess this,
we have first analyzed the MECS X-ray spectrum obtained from 1E~1743.1-2843, 
which is located at $13\arcmin$ from SAX~J1747.0-2853. This 
is rather soft and satisfactorily described by a single 
blackbody component with kT=($2.10\pm0.08$)~keV 
(our fit yields a reduced $\chi^{2}_{r}$=1.14 for 156 dof). The 
total 2-10~keV intensity (unabsorbed) is $\sim10$~mCrab, consistent
with other NFI measurements (\cite{Cre99}). The 
contribution of 1E~1743.1-2843 is $\sim15$~\% of the total energy flux 
detected by the High Pressure Gas Scintillation Proportional Counter (HPGSPC) 
in the range 4-25~keV. We then decided not to include 
the HPGSPC data in our analysis.

Additional contamination at high energies may come 
from other sources outside the FOV of the concentrators. 
Among these are A1742-294, a persistent though variable X-ray burster
and the microquasar 1E~1740.7-2942,
which are $0.68\arcdeg$ and $1.16\arcdeg$ off the pointing position, 
respectively. 
The intensity from A1742-294 was $\sim30$~mCrab in the 2-28~keV energy 
band as detected by the WFCs on March 21-22, but the extrapolation of its 
thermal spectrum (\cite{Pav94}) to high 
energies implies only a $\leq1$\% contamination above $\sim30$~keV.
1E~1740.7-2942 has been extensively monitored by 
{\em RXTE}~PCA instrument (\cite{Mai99}). Its intensity on 1998, March~23 was
$\sim35$~mCrab (2-10~keV). This source features a very hard tail, and on 
the basis of the
hardness ratio usually observed we can estimate its intensity 
being $\sim50$~mCrab in the 30-150~keV energy band. Hence its expected 
contribution to the detected flux in the PDS is  
$\leq10$~\% in the whole energy range of the instrument. 
Activity from the soft transient source KS1741-29, located at 
$\approx0.7$ degrees from SAX~J1747.0-2853 was also detected on 
March 12 (\cite{inz99}) with a 2-10~keV flux of $\sim9$~mCrab. The source
then declined to $\sim2$~mCrab on March 31, and a soft spectrum was
measured (\cite{Sid99}). The extrapolation of this spectrum to high 
energies yields an expected contamination of $\sim3-4$\% for the PDS
"on source" data at 30 keV. 

Conversely, we note that any contamination in the PDS from a
source having an emission spectrum similar to SAX~J1747.0-2853
would be negligible as far as we apply the standard 
procedure of leaving free the PDS normalization  
relative to MECS, which is done in order to accommodate 
cross-calibration uncertainties (see e.g., \cite{Gua98}).  
The additional systematic error in the spectral channels is then 
due only to possible differences in spectral shapes. 
On the basis of the previous assumptions, we estimate this to
be less than $\sim5$\% whereas the overall uncertainty in the high energy
flux should be of the order of 10\%. 

The bandpasses for the spectral analysis were limited to 1.0-3.0~keV for the 
LECS 
and 1.8-10.5~keV for the MECS to take advantage of more accurate detector 
calibrations. LECS data below 1~keV were ignored due to the high
absorption and low source flux. The PDS data analysis was limited
to energies above 32 keV in order to avoid as much as possible contamination.

\begin{figure}
\plotone{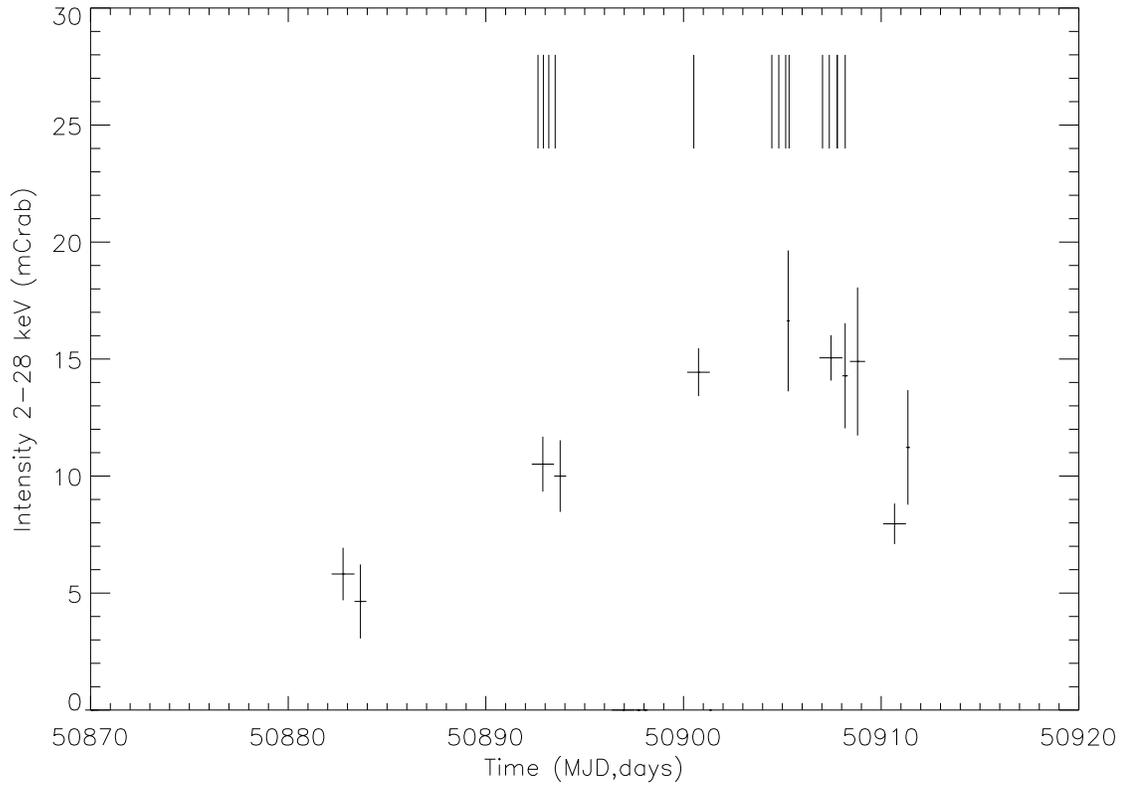}
\caption{
    WFC light curve of the X-ray outburst of SAX~J1747.0-2853 in 
    spring 1998. The markers indicate the epoch of the observed
    bursts.
   \label{fig1}}
\end{figure}

\placefigure{fig1}

% ------------------- Section 3: The wide band persistent emission ----------------- 
\section{THE WIDE BAND PERSISTENT EMISSION}

The WFC-measured outburst light curve of SAX J1747.0-2853 is 
shown in Fig.~1 (see \cite{Jag97} for details on WFC data reduction
and analysis). The source 
reached a peak intensity of $\sim15$~mCrab with a rather slow rise and 
apparently rapid decay. 
SAX J1747.0-2853 was observed by {\em BeppoSAX}~NFI about 10 days
before the outburst maximum. The source intensity
is found to increase by $\sim20$~\% in 20 hours during this observation.

We have analysed the wide band NFI spectrum of the averaged emission in the
energy range 1-200~keV. A hard X-ray tail is present (as detected
by the PDS up to $\sim150$~keV) which can be described by a power law 
with photon spectral index $\Gamma=2.4\pm0.1$ and a corresponding intensity 
of $\sim5\times10^{-10}$~erg~cm$^{-2}$~s$^{-1}$ in the 30-150 keV range.  
The absorption column is relatively high
(${N}_{H}$~$\simeq$~10$^{23}$~cm$^{-2}$),
consistent with the measured values of the column 
density in the spectra of sources observed in the Galactic Centre (GC) 
region (\cite{Sid99}). Modelling the whole X-ray spectrum with an absorbed
(\cite{Mor83}) power law or an 
exponentially cutoff power law yields to very poor fits (see Table 1 for
details), leaving a large   
residual structure near $\approx7$~keV. This 
can be interpreted either as a broad iron line feature or the lower 
energy part of an absorption edge. We tried to fit the observed feature 
adding a broad 'iron like' line to the continuum, using both exponentially
cutoff power law and thermal Comptonization (\cite{Tit94}). In all cases 
the fit results imply unacceptably high values of the
PDS/MECS normalization. 

Modelling the feature with a sharp absorption edge near 7 keV
provides an appropriate value of the normalization, but the reduced 
$\chi^2$ of 1.38 is still too high. Conversely, a relatively good fit can 
be obtained for thermal Comptonization by including an additional soft 
component ($\chi^2$=1.1), yielding an absorption edge at $\approx7.5$~keV. 

The details of the fits are given 
in Table 1. The soft component may be described by blackbody emission with
color temperature of $\sim0.6$~keV, or by a multi-color disk 
blackbody (\cite{Mit84}), with
a temperature ${kT}_{in}$~$\sim0.8$~keV for the inner disk region. The
significance of the soft component is high if one considers the improvement 
in $\chi^2$ obtained for the Comptonization continuum (from 1.8 to 1.1)
whereas no acceptable fit is obtained for the exponentially cutoff power 
law. A visual inspection of the residuals near 7 keV led us to search for
an additional emission line component. Using  
a gaussian line profile yields a narrow line emission with energy 
${E}_{K\alpha}$=($6.90\pm0.09$)~keV and an equivalent width of ($55\pm14$)~eV,
with a reduced $\chi^2$ of 0.97 over 74 dof.

The significance of this detection is high (the F-test provides
a value of 9.9 of the F-statistic, for a null hypothesis probability of
$1.4\times10^{-5}$). We caution however, about
the presence in this region of a hot thermal plasma with strong line
emission from both neutral and ionized species of iron (\cite{Koy96}). 
In fact,
the detected line could be partly originated by the plasma emission in
the line of sight of SAX~J1747.0-2853, or could arise as a systematic
effect as far as the intensity of the
diffuse component is variable on a spatial scale 
smaller than our source and background extraction regions.  
 
The unfolded energy spectrum obtained for the blackbody plus Comptonization 
fit and narrow line is shown in Fig.~2.

\begin{figure}
\plotone{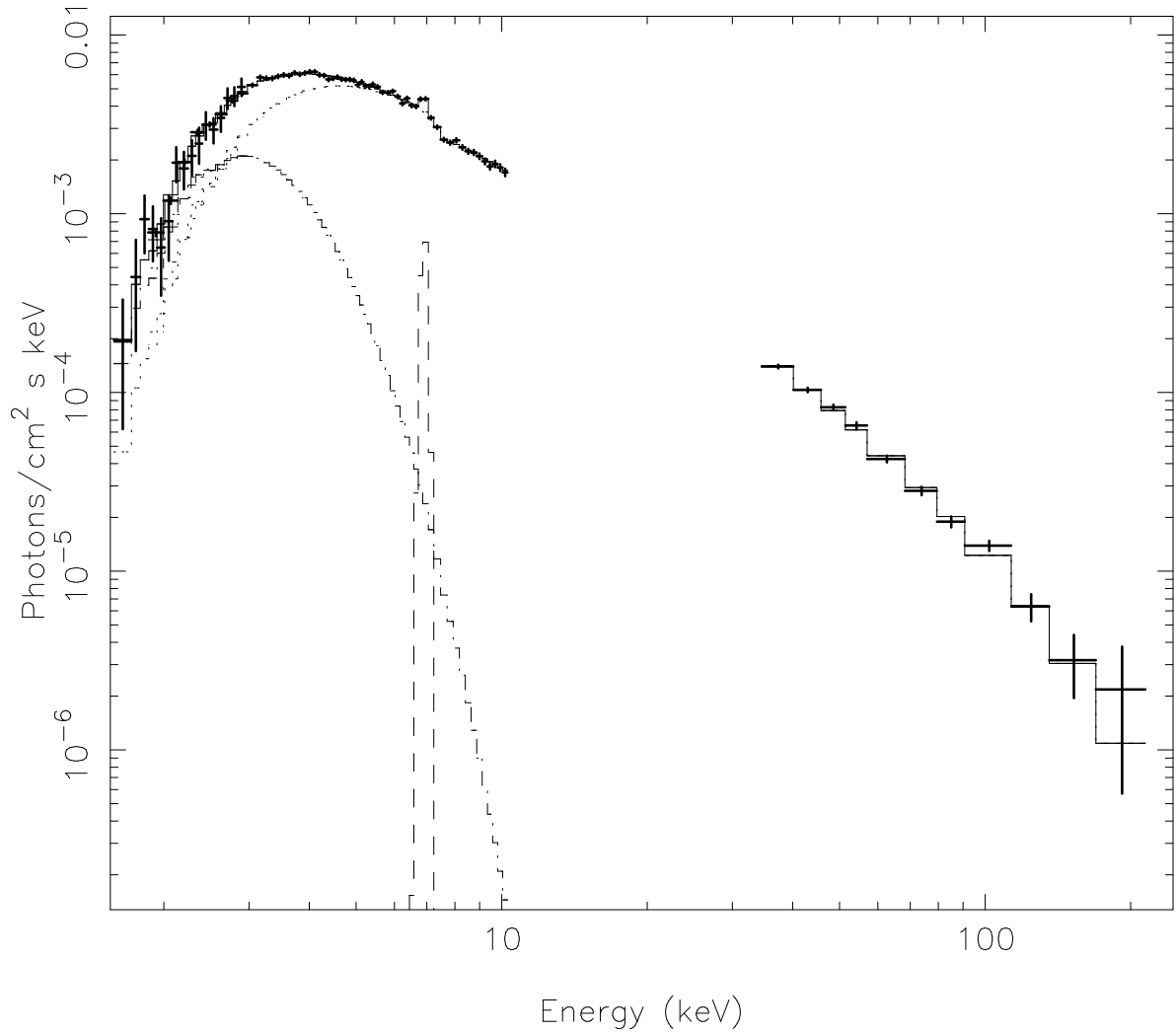}
\caption{
The unfolded broadband spectrum of SAX~J1747.0-2853 as measured by the
{\em BeppoSAX} NFIs on March 23-24, 1998 during a 72 ksec exposure.
The model fit to a blackbody (dotted curve) plus 
thermal Comptonization (dot-dashed curve) with absorption edge and 
iron line is shown. 
\label{fig2}}
\end{figure}

\placefigure{fig2}

% ------------------- Section 4: X-RAY BURSTS  ----------------- 
\section{X-RAY BURSTS}

In 1998, between MJD~50892.6 and MJD~50911.4 a total of 14 X-ray bursts
was observed from SAX~J1747.0-2853. 
All these events show similar time and
spectral characteristics, with fast rise, exponential decay and 
spectral softening. 
One rather strong event occurred on MJD~50900 with a peak intensity of   
$(820\pm70)$~mCrab (2-28~keV). For this burst, 
the time profiles in two different energy bands
(2-8~keV, 8-28~keV) are both
characterized by fast rise times ($\leq2$~s) and longer exponential 
decays with e-folding times depending on energy band (see Fig.~3). 
In particular, the high energy time history shows clear evidence for 
double-peaked profile and its  
e-folding time is evidently shorter ($\sim3$~s) than the 
one of the low energy profile ($\sim8$~s).
The average burst spectrum is consistent with absorbed blackbody
emission with average colour temperature of $\sim2$~keV and the average 
blackbody radius of the hypothetical emitting sphere is $\sim7$~km 
(assuming 10~kpc distance). The absorption ${N}_{H}$ 
was set to $10^{23}$~cm$^{-2}$ as derived from the persistent 
emission observed by the NFI.  

Four time-resolved spectra were accumulated in order to study
the time evolution of the spectral parameters. The
spectra were subtracted for the source persistent emission, and their time
intervals were chosen to match the first-peak, interpeak,
second-peak and decay phases in the 8-28~keV burst time
profile. Since 
the obtained emission spectra are all consistent with blackbody,
we could determine the time evolution of the radius of the emitting sphere 
(see lower panel of Fig.~3).
The blackbody radii were calculated assuming isotropic emission at a
source distance of 10~kpc and not correcting for gravitational redshift
and conversion to effective blackbody temperature from colour temperature. 
A radius expansion by a factor of $\sim2$ is observed.                    

\begin{figure}
\plotone{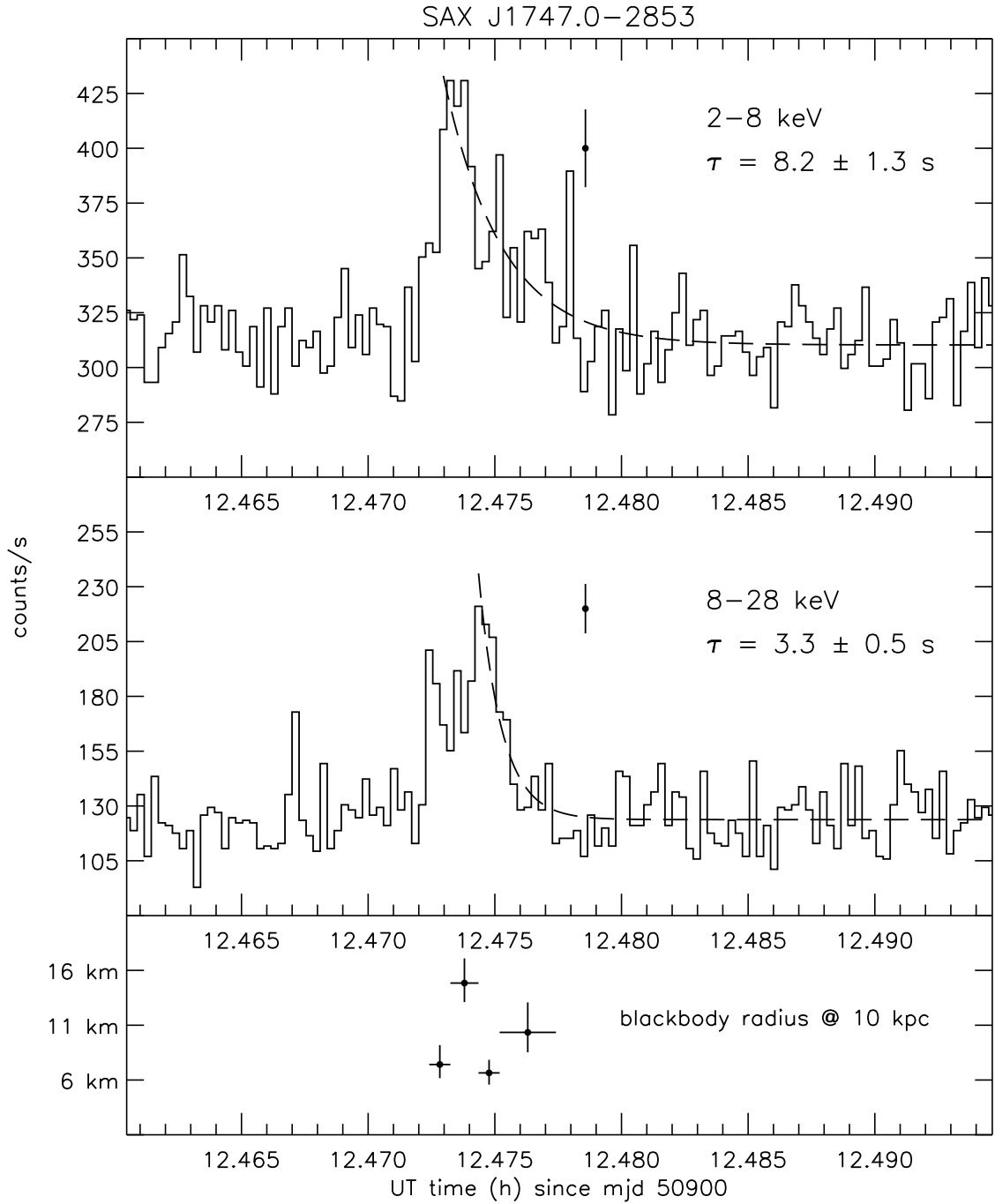}
\caption{
    Time profile in the low (2-10 keV, top panel) and high
    (10-28 keV, middle panel) energy range of the MJD~50900
    burst. 
    The time evolution of
    the blackbody radius, normalized to a distance of 10 kpc, is also shown.
    The actual blackbody radius is a linear function of source 
    distance.  
\label{fig3}}
\end{figure}

\placefigure{fig3}

% ------------------- Section 5: CONCLUSIONS	 ----------------- 
\section{CONCLUSIONS}

On the basis of the spectral and
timing properties of the first bursts observed by the WFCs it is clear 
that the transient source 
SAX~J1747.0-2853 is a type~I X-ray burster, thus indicating  
that the compact object is a neutron star and that the source should be 
classified as a candidate low-mass X-ray binary (LMXB). 

The photospheric expansion observed for 
the MJD~50900 burst can be interpreted as adiabatic expansion
during a high luminosity, Eddington-limited type~I burst.
Such bursts may be used
to estimate the source distance (assuming isotropic burst emission). 
Given the observed blackbody spectrum, the burst peak intensity corresponds 
to an unabsorbed bolometric flux of 
$(3.18\pm0.27)\times10^{-8}$~erg~cm$^{-2}$~s$^{-1}$. Using the standard 
candle luminosity of $(3.0\pm0.6)\times10^{38}$~erg~s$^{-1}$,
empirically calculated by \cite{Lew95} on a sample of 
Eddington limited bursters, leads to a  
distance value $d=(8.9\pm1.3)$~kpc for SAX~J1747.0-2853. On the other
hand, assuming the theoretical Eddington luminosity for a typical
1.4~${M}_{\odot}$~NS ($2.5\times10^{38}$~erg~s$^{-1}$) and burst isotropy, 
we obtain $d=(8.1\pm0.3)$~kpc (the error is purely statistical).
The calculated distance of $\sim9$~kpc implies an average radius of the  
blackbody emitting region of $\sim6$~km as obtained from the observed colour 
temperature, i.e. without correcting for 
colour-to-effective temperature ratio which could lead to an underestimate
of the observed radius by a factor $\sim2$ (\cite{Ebi87}).

The wide band NFI spectrum of SAX~J1747.0-2853 obtained on 
1998, March 23 (when the source was close to outburst maximum) 
can be explained as
being composed of a primary non-thermal component (detected up to 
$\sim150$~keV) which can be fitted by Comptonization of soft 
photons at $\sim1.3$~keV, plus a soft thermal component compatible with 
blackbody emission at $\sim0.6$~keV or 
disk blackbody with an inner temperature of $\sim0.8$~keV.
The observed spectral shape (Fig.~2) suggests that SAX~J1747.0-2853 
was in a typical low/hard state during the 1998 outburst, with a luminosity
ratio L/${L}_{Edd}\simeq0.025$. The presence of
the soft component, with a total flux of 
$\sim3\times10^{-11}$~erg~cm$^{-2}$~s$^{-1}$ (i.e., $\sim10$\% of the 
total 2-10~keV flux) is in line 
with recent detections of a soft excess in the spectra of
a number of type~I bursters observed in hard state 
(e.g., \cite{Bar00}; \cite{Nat00}) with  
temperatures typically in the range 0.5-2~keV. 

The spectrum shows evidence of strong reprocessing, suggestive of
Compton reflection and/or absorption of primary X-ray photons on
optically thick material (\cite{Lig88}). This could be an accretion disk 
around a central hard X-ray source consisting of a hot plasma 
located either in the inner disk or in a NS boundary layer (\cite{Pop00}).
The reprocessing of the hard X-rays is clearly revealed by the 
presence of an iron feature, which can be described as a combination
of a rather sharp absorption edge at nearly 7.4~keV and a weak, narrow 
Fe~K$\alpha$ emission line at $\approx6.9$~keV with an equivalent width 
of 55~eV. 
The energies of the edge and of the narrow line (see Section~3) seem to 
suggest that the material is ionized rather than cold 
%(Ross, Fabian \& Young 1999).
(\cite{Ros99}).
The presence of ionized material even at low accretion rate can be
expected as far as the effective temperature of the inner 
disk is of the order of 0.5~keV (\cite{Ros93}).  

Detections of iron edge features are generally  
accompanied by a characteristic hardening of the spectrum above 10~keV
due to the increase of the scattering opacity at these energies 
(Compton reflection). In our case the amount of the reflection component 
is difficult to assess, since we do not have useful data in the range 
10.5-32~keV and also due to the uncertain behaviour of the relative 
PDS/MECS spectra normalization (see discussion in Section~2).

\acknowledgments

We thank Team Members of the BeppoSax Science Operation Centre and Science 
Data Centre for continuos support  
and timely actions for near-real-time detection 
of new transient and bursting sources and the follow-up TOO  
observations. The {\em BeppoSAX} satellite is a joint Italian and 
Dutch programme.

\clearpage
\begin{deluxetable}{llll}
%\rotate
\scriptsize
\tablecaption{Results of spectral model fits to the persistent
emission in the energy range 1-200 keV}
\label{tbl1}
%\tablewidth{0pt}
\tablecolumns{4}
\tablehead{
         \colhead{Model\tablenotemark{a}}
       & \colhead{$\rm N_{H}$\tablenotemark{a}}
       & \colhead{Model parameters\tablenotemark{b}}
       & \colhead{$\chi^2_\nu$\tablenotemark{c}}
}
\startdata
Powerlaw                                 & $9.4\pm0.2$ &
 $\Gamma=1.95\pm0.03$ &
 4.45[82] \\
Cutoffpl                                 & $8.9\pm0.2$ &
 $\Gamma=1.80\pm0.03$, ${E}_{c}$=$101.2\pm0.2$ &
 2.88[81] \\
Edge(Cutoffpl)                           & $8.0\pm0.2$ &
 $\Gamma=2.06\pm0.03$, ${E}_{c}$=$72.3\pm0.1$, ${E}_{edge}$=$7.56\pm0.05$ &
 1.38[79] \\
Edge(BB+Cutoffpl)\tablenotemark{d}                & $9.9\pm0.6$ &
  $\Gamma=1.68\pm0.05$, ${E}_{c}$=$85.1\pm0.1$, ${kT}_{bb}$=$0.34\pm0.02$ &
 1.35[77] \\
 &  &
${R^2}_{bb}$=$2189^{+5940}_{-1610}$, ${E}_{edge}$=$7.57\pm0.07$ & \\
Edge(Comptt)                             & $9.9\pm0.6$ &
 ${kT}_{0}$=$0.98\pm0.03$, ${kT}_{e}$=$23\pm2$, $\tau=4.6\pm0.4$ &
 1.81[78] \\                              
Edge(BB+Comptt)\tablenotemark{d}           & $7.6\pm0.6$ &
 ${kT}_{0}$=$1.3\pm0.2$, ${kT}_{e}$=$35\pm12$, $\tau=2.7\pm1.1$,
 ${kT}_{bb}$=$0.58\pm0.06$  &
 1.10[77] \\
 &  &
${R^2}_{bb}$=$147^{+220}_{-77}$, ${E}_{edge}$=$7.47\pm0.06$ & \\       
Edge(BB+Comptt+Line)\tablenotemark{d}           & $8.0\pm0.9$ &
 ${kT}_{0}$=$1.1\pm0.2$, ${kT}_{e}$=$31\pm7$, $\tau=3.3\pm0.8$,
 ${kT}_{bb}$=$0.55\pm0.10$  &
 0.97[74] \\
 &  &
 ${R^2}_{bb}$=$137^{+369}_{-75}$, ${E}_{edge}$=$7.41\pm0.16$,
 ${E}_{K\alpha}$=$6.90\pm0.09$ ($EW\sim55$ eV)   &  \\
Edge(MCD+Comptt+Line)\tablenotemark{d}         & $7.9\pm0.7$ &
 ${kT}_{0}$=$1.3\pm0.3$, ${kT}_{e}$=$33\pm9$, $\tau=3.1\pm0.9$,
 ${kT}_{in}$=$0.82\pm0.23$  &
 0.97[74] \\
 &  &
 ${R^2}_{in}$~$\cos{\theta}$~=~$80^{+265}_{-64}$, 
 ${E}_{edge}$=$7.44\pm0.10$ & \\
 &  &
 ${E}_{K\alpha}$=$6.89\pm0.07$ ($EW\sim55$ eV)   &  \\
 
\enddata
\tablenotetext{a}{Follows the same notation as in XSPEC v.11. Wisconsin
                  absorption is used in all models.}
\tablenotetext{b}{Parameters description and units: \\
$\rm N_{H}$, Wisconsin absorption parameter in units of
 $10^{22}$~cm$^{-2}$;
                 $\Gamma$, photon index of power law component;
                 ${E}_{c}$, high energy cutoff value, in keV;
                 ${kT}_{0}$, temperature of the comptonized soft seed
photons, in keV;
                 ${kT}_{e}$, temperature of the electron plasma,in keV;
                 $\tau$, plasma optical depth for spherical geometry;
                 ${kT}_{bb}$, temperature of the blackbody component, in
keV;
                 ${R^2}_{bb}$, normalization expressed as $r^2/d^2$,
where r is the blackbody radius in km and
d is the distance to the source
in units of 10 kpc;                                           
                 ${R^2}_{in}$~$\cos{\theta}$, where ${\theta}$ is the
disk inclination angle and ${R}_{in}$ is the disk inner radius in km,
for a source at 10 kpc;
                 ${E}_{edge}$, energy of absorption edge, in keV;
                 ${E}_{K\alpha}$, centroid energy of gaussian line
profile, in keV
                 }
\tablenotetext{c}{Number of d.o.f. is given in parentheses}
\tablenotetext{d}{Inner and upper bounds for the soft component normalization
correspond to 90\% confidence level}

\tablecomments{Errors are single parameter l$\sigma$ errors.}
\end{deluxetable}

\clearpage
      
\clearpage
 
\clearpage
 
\end{document}